\documentclass[runningheads]{llncs}
\usepackage{butterma}
\idline{The final publication is available at
\href{http://www.springerlink.com/content/m644176548m673q8/}{www.springerlink.com}\\
S.~Bechhofer et al.\ (Eds.): ESWC 2008, LNCS 5021}
\renewcommand{\year}{2008}

\newif\ifdraft
\draftfalse

\usepackage[T1]{fontenc}
\usepackage[utf8]{inputenc}
\usepackage{textcomp}

\ifdraft
\usepackage[show]{ed}
\usepackage{pdfsync}
\else
\fi

\usepackage{tikz}

\def\thetitle{SWiM -- A Semantic Wiki for Mathematical Knowledge Management}

\definecolor{NavyBlue}{cmyk}{0.94,0.54,0,0.3}
\usepackage[pdfpagescrop={91 71 521 721},a4paper=false,pdftex,pdfstartview=FitV,plainpages=false,pdfpagelabels,colorlinks=true,linkcolor=NavyBlue,citecolor=NavyBlue,urlcolor=NavyBlue,hypertexnames=true]{hyperref}
\hypersetup{
    pdfauthor = {Christoph Lange},
    pdftitle = {\thetitle},
    pdfkeywords = {Semantic Wiki OMDoc Ontology Services Science Mathematical
Knowledge Management Mathematics}
}
\usepackage{url}

\usetikzlibrary{arrows}

\title{\thetitle} \author{Christoph Lange} \institute{Computer Science, Jacobs University
  Bremen, \email{ch.lange@jacobs-university.de}}

\begin{document}

\maketitle
\thispagestyle{electronic}

\begin{abstract}
  SWiM is a semantic wiki for collaboratively building, editing and browsing mathematical
  knowledge represented in the domain-specific structural semantic markup language OMDoc.
  It motivates users to contribute to collections of mathematical knowledge by instantly
  sharing the benefits of knowledge-powered services with them.  SWiM is currently being
  used for authoring content dictionaries, i.\,e.\ collections of uniquely identified
  mathematical symbols, and prepared for managing a large-scale proof formalisation
  effort.
\end{abstract}

\section{Research Background and Application Context: Mathematical Knowledge Management}
\label{sec:background}

A great deal of scientific work consists of collaboratively authoring
\emph{documents}---taking down first hypotheses, commenting on results of experiments,
circulating informal drafts inside a working group, and structuring, annotating, or
reorganising existing items of knowledge, finally leading to the publication of a
well-structured article or book.  Here, we particularly focus on the domain of mathematics
and on tools that support collaborative authoring by utilising the knowledge contained in
the documents.  In recent years, several \emph{semantic markup} languages have been
developed to represent the clearly defined and hierarchical structures of mathematics.
The XML languages MathML~\cite{CarlisleEd:MathML07}, OpenMath~\cite{BusCapCar:2oms04}, and
OMDoc~\cite{Kohlhase:omdoc1.2} particularly aim at exchanging mathematical knowledge on
the web.  OMDoc, employing Content MathML or OpenMath representing the functional
structure of mathematical \emph{formul\ae}---as opposed to their visual appearance---and
adding support for mathematical \emph{statements} (like symbol declarations or axioms) and
\emph{theories}, has many applications in publishing, education, research, and data
exchange~\cite[chap.~26]{Kohlhase:omdoc1.2}.  The main challenge is \emph{acquiring} a
large collection of OMDoc-formalised knowledge that can power such added-value services.
In an open, collaborative environment, the workload can be distributed among many authors,
but as semantic markup makes fine-grained structures explicit, it is tedious to author.
As the community can only benefit from added-value services after a substantial initial
investment (writing, annotating and linking) on the author's part, we sought for
motivating authors into action by offering ``elaborate […] services for the concrete
situation'' they are in~\cite{KohMue:added-value07}.

\section{Key Technology: Semantic Wiki and Ontologies}
\label{sec:semwiki}

Our research is motivated by the assumption that in this context a semantic wiki comes in
handy.  OMDoc supports all levels of formalisation, from human-readable texts to fully
formal representations for automated theorem proving, and semantic wikis have been found
appropriate for collaboratively refining knowledge models (cf.\
\cite{Schaffert:SemanticSocialSoftware06}).  User motivation in semantic wikis by instant
gratification has been investigated in earlier works~\cite{aumueller05:wiksar}.  The
ultimate goal of our work is to achieve a feedback loop where users are supported to
contribute well-structured knowledge, which is then exploited to offer services, which in
turn facilitate editing and motivate new contributions~\cite{Lange:SWiMSciColl07}.

\begin{figure}
  \centering
  \vspace{-.065cm}
  \begin{tikzpicture}
    \node[text width=3.7cm,draw] (page) at (0,0) {<omdoc>\\
~~<proof id="pyth-proof"\\
~~~~for="pythagoras">\\
~~~~\ldots</proof>\\
</omdoc>};
    \node (rdfdummy) at (4,0) {};
    \draw[->,dashed] (page) -- node [below] {extraction} node[above] {RDF} (rdfdummy);
\begin{scope}[xshift=5cm,xscale=3,yscale=.8,>=latex,font=\scriptsize]
  \tikzstyle abox=[font=\scriptsize,draw,minimum height=2.5ex,rounded corners];
  \tikzstyle tbox=[font=\scriptsize\bfseries\itshape,draw,minimum height=2.5ex];
  \node[abox] (pp) at (0,0) {pyth-proof};
  \node[abox] (pt) at (1,0) {pythagoras};
  \node[tbox] (P) at (0,1) {\bfseries Proof};
  \node[tbox] (T) at (1,1) {\bfseries Theorem};
  \draw[-open triangle 60] (pp) -- node[left=.4em] {type} (P);
  \draw[-open triangle 60] (pt) -- node[right=.5em] {type} (T);
  \draw[->] (pp) -- node[below=-.3ex] {proves} (pt);
  \draw[->] (P) -- node[below=-.3ex,font=\scriptsize\bfseries\itshape] {proves} (T);
\end{scope}

    \node[text width=5.25cm,draw,anchor=west] (triples) at (4,-.8) {\scriptsize
<pyth-proof, rdf:type, omdoc:Proof>\\
<pyth-proof, omdoc:proves, pythagoras>};
  \end{tikzpicture}
  \caption{RDF extraction from OMDoc markup in a wiki page}
  \label{fig:rdf-extraction}
  \vspace{-.065cm}
\end{figure}
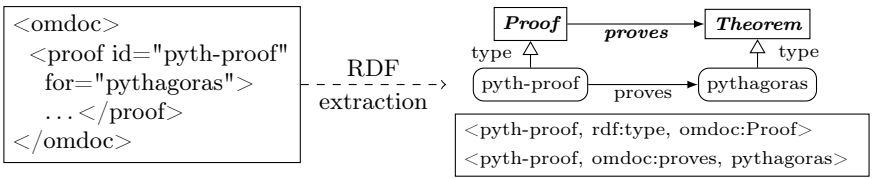

Semantic markup has deep structures: an OMDoc document can contain theories containing
statements that contain formul\ae\ referring to symbols defined in other theories.  This
is uncommon for most semantic wikis, where the structures are rather flat and one aims at
small pages to prevent editing conflicts and to facilitate search and navigation.  So to
adapt OMDoc's model of knowledge to a semantic wiki, we had to choose an appropriate
granularity of wiki pages and arrived at one page holding one mathematical statement or
one theory.  To make knowledge from OMDoc documents usable on the semantic web,
information about the resources represented by pages and their interrelations (e.\,g.\ ``a
\emph{proof} \emph{for} the Pythagorean theorem'') are extracted to RDF.  As a vocabulary
for this, we modeled OMDoc's structures explicitly in a \emph{document
  ontology}~\cite{Lange:SWiMSciColl07} in OWL-DL.  This ontology contains e.\,g.\ the
information that both theorems and proofs are specialisations of a general ``mathematical
statement'', and that a proof can prove a theorem (Fig.~\ref{fig:rdf-extraction}).
Moreover, generic transitive dependency and containment relations have been modeled.  For
example, having one theory import another theory (and reusing symbols defined there)
establishes a dependency.  One theory logically contains its statements; similarly,
statements can contain sub-statements, as in the case of a proof that consists of multiple
steps.

\section{The SWiM 0.2 Prototype: IkeWiki + OMDoc}
\label{sec:swim}

As a base system for the implementation, we chose
IkeWiki~\cite{schaffert06:STICA-ikewiki}.  Among the systems evaluated, it offered the
richest XML infrastructure---a key requirement for adding OMDoc support---and was found to
be most extensible~\cite{Lange:swmkm-tr07}.  Its backend consists of a PostgreSQL database
for the page contents, a Jena RDF store for the RDF graph and the ontologies.  Additional
ontologies can easily be imported.  The frontend heavily relies on the Dojo Ajax toolkit.

Technically, the extension of IkeWiki to SWiM required supporting OMDoc in addition to the
HTML-like wiki page format.  To foster stepwise formalisation of informal text, we chose
to mix OMDoc fragments with wiki markup.  Thus we could still rely on IkeWiki's WYSIWYG
HTML editor, which just had to be enhanced by support for OMDoc XML elements.  Moreover,
this choice allowed for an easier maintenance of the OMDoc-related enhancements to the
SWiM code base and avoided changes to the underlying database schema.  The document
ontology is preloaded into the RDF store.  RDF triples are extracted from the OMDoc markup
upon saving a page or importing an OMDoc file.  Additional XSLT template rules care for
rendering embedded OMDoc fragments.  In order to render mathematical formul\ae, there is a
\emph{notation definition} for every semantic symbol.  These notation definitions can be
imported and edited right in the wiki, as parts of OMDoc
documents~\cite{lange:swim-notation-semantics08}.  An efficient, specialised renderer
supporting the upcoming MathML 3 standard~\cite{mmlproc:web,CarlisleEd:MathML07} applies
them to the symbols in the formul\ae.  In the editing view, statement- and theory-level
structures of OMDoc are made accessible as special HTML tables, whereas mathematical
formul\ae\ given in semantic markup are made accessible in a simplified ASCII notation of
OpenMath.  OMDoc documents are browsable via inline links manually set in the informal
parts, via links from occurrences of symbols in formul\ae\ to the place of their
declaration, set by the formula renderer, and via RDF links, displayed in a separate box
by IkeWiki.  The latter comprise those triples that are extracted from the markup (cf.\
Fig.~\ref{fig:rdf-extraction}), as well as triples inferred by a reasoner\footnote{The
  ontology is prepared for DL reasoning, but currently only the RDFS reasoner built into
  Jena is used.}.

\begin{figure}
  \centering
  \includegraphics[width=.9\textwidth]{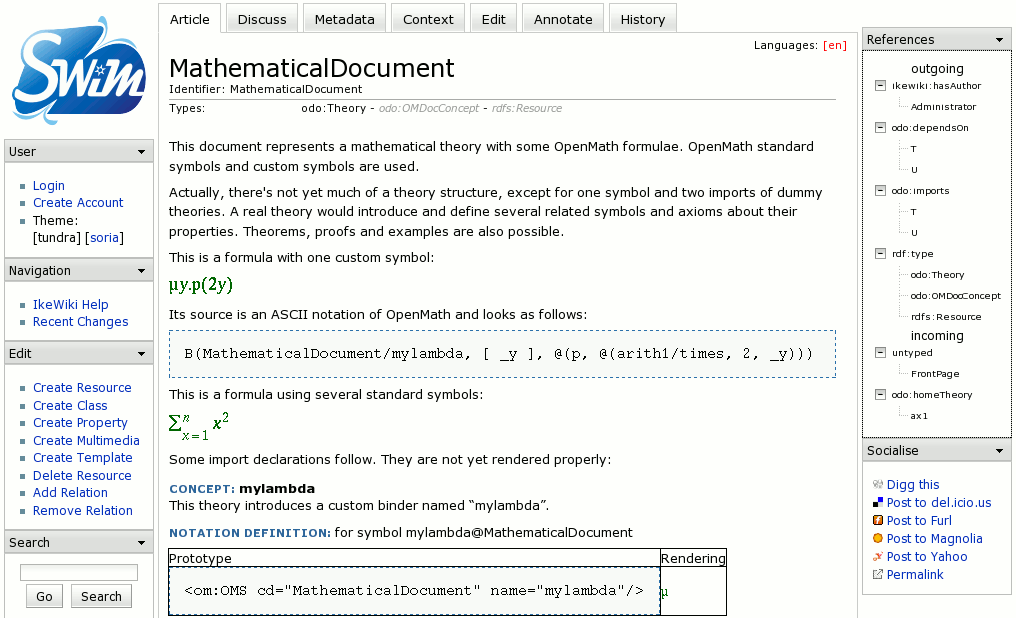}
  \caption{A mathematical document in SWiM}
  \label{fig:swim-page}
\end{figure}

SWiM also relies on the ontology for reacting on changes to notation definitions.  When an
author changes a notation definition $n$ for a symbol $s$, exactly those wiki pages that
contain a formula using $s$ or that include other pages containing such formul\ae\ need to
be re-rendered.  Looking up the symbol $s$ rendered by $n$, the formul\ae\ $f_i$ using $s$,
or pages (transitively) including the $f_i$ would be clumsy in the OMDoc XML sources, but
is easy in the RDF graph, as this information is extracted from the documents and
represented using ontology properties such as \textit{NotationDefinition--renders--Symbol}
and \textit{Statement--contains--Formula; Formula--uses--Symbol}.  This service allows for
instant visual debugging of notation definitions~\cite{lange:swim-notation-semantics08}.
For upcoming releases, more ontology-powered services are planned, including more general
change management, learning assistance, and editing facilitations like editing of
subsections and auto-completion of link targets~\cite{swim-roadmap}.  There is some
evidence that many services can be based on the most generic relations of dependency and
(physical or logical) containment~\cite{Lange:SWiMSciColl07}.  With scientists and
knowledge engineers in mind, we envisage SWiM as a development environment that
conveniently supports refactorings of knowledge\footnote{This is common in mathematics,
  e.\,g.\ in algebra: If one just needs groups, they can be defined by a theory with the
  four well-known axioms.  For explicitly modeling related structures as well, one would
  break this into smaller theories---\emph{semigroup} just defining an associative
  operation on a set, \emph{monoid} importing this and extending it by an identity
  element, and finally the refactored \emph{group}, adding inverse elements.}.

\section{Use Cases and Applications}
\label{sec:usecase}

Now that viewing, browsing, editing, importing and exporting mathematical documents
basically works, we are evaluating SWiM in practical settings.  The \textbf{Flyspeck}
project is about large-scale formalisation of a proof of the Kepler conjecture.  We are
starting to support this effort by ``crowdsourcing'' the knowledge compiled so far
(hundreds of proof sketches that are not yet machine-verifiable) on a SWiM
site~\cite{LangeMcLRabe:FlyspeckWiki08}.  The main challenge is giving an interested
visitor an impression of the extent of the project and, using appropriate SPARQL queries,
showing him where work needs to be done.  Currently we are investigating how the original
{\LaTeX} sources can be utilised by automatically converting them to HTML with MathML,
then to informal OMDoc, breaking that into wiki pages, and letting the users formalise
them stepwisely.  For the upcoming \textbf{OpenMath~3} standard, SWiM is currently being
extended to an editor for OpenMath Content
Dictionaries~\cite{lange:swim-notation-semantics08}, which could be regarded as flat OMDoc
theories that just define symbols and do not import anything.  There, mainly editing
Dublin Core metadata and notation definitions is of interest.

\section{Conclusion and Related Work}
\label{sec:conclusion}

SWiM makes mathematical documents editable collaboratively and particularly facilitates
browsing them by exploiting the knowledge they contain.  Domain-specific services are
powered by an ontology that models structures of documents---an advantage over generic
semantic wikis, which would not be able to offer additional services for mathematical
knowledge.  Competing non-semantic approaches like the math encyclop\ae dia
\textsl{PlanetMath} (evaluated in~\cite{Lange:swmkm-tr07}) are less flexible, as they
cannot exploit the structures of their presentation-oriented \LaTeX\ formul\ae\ and rely
on a fixed set of metadata.  Most services for editing and browsing need to be hard-coded,
which potentially restricts the scale of knowledge managment tasks the systems can be
applied to.  The SWiM approach of integrating a semantic markup language into a wiki by
choosing an appropriate page granularity, modeling a document ontology, and extracting
relevant facts from the markup into RDF has successfully been applied to OMDoc and the
closely related but syntactically different
OpenMath~\cite{lange:swim-notation-semantics08} and is likely to be portable to other
domains as well, e.\,g.\ for the chemical markup language CML.

\bibliographystyle{abbrv}
\bibliography{swim-demo}


\ifdraft
\ednotemessage
\fi
\end{document}